# Classification of time-domain waveforms using a speckle-based optical reservoir computer

Uttam Paudel,[†] Marta Luengo-Kovac, Jacob Pilawa, T. Justin Shaw, George C. Valley[*]

*Photonics Technology Department—The Aerospace Corporation, 2310 E El Segundo Blvd, El Segundo, CA, 90245*
*[†] uttam.paudel@aero.org*
*[*]george.c.valley@aero.org*

**Reservoir computing is a recurrent machine learning framework that expands the dimensionality of a problem by mapping an input signal into a higher-dimension reservoir space that can capture and predict features of complex, non-linear temporal dynamics. Here, we report on a bulk optical demonstration of an analog reservoir computer using speckles generated by propagating a laser beam modulated with a spatial light modulator through a multimode waveguide. We demonstrate that the hardware can successfully perform a multivariate audio classification task performed using the Japanese vowel speakers public data set. We perform full wave optical calculations of this architecture implemented in a chip-scale platform using an $SiO_2$ waveguide and demonstrate that it performs as well as a fully numerical implementation of reservoir computing. As all the optical components used in the experiment can be fabricated using a commercial photonic integrated circuit foundry, our result demonstrates a framework for building a scalable, chip-scale, reservoir computer capable of performing optical signal processing.**

## Introduction

The rapid growth of data traffic leading to the emerging big data era has made special-purpose signal processing hardware a necessary tool. In particular, an analog processor capable of performing on-the-fly signal processing would significantly reduce the overhead of the storage and transmission of large datasets and the need for post-processing [1]. Over the past few years, several realizations of optical signal processors [2,3], such as optical neural networks [4-6] and optical reservoir computers [7-10], have been demonstrated that generate random matrices and perform matrix multiplication passively at "no cost" at the speed of light. Electrical neural networks require the storage of large matrices of weight coefficients, which inherently slows down the computation as the information must be passed back and forth between the memory and processing units and is energy costly. In an optical neural network, the weight coefficients are an intrinsic part of the photonic system, and the information is transmitted at the speed of light [11].

Photonic reservoir computing (RC), in particular, has attracted a lot of attention [2,3], as it can be used for a wide variety of classification, prediction, and memory tasks [8-21]. Due to its recurrent nature, it can mimic and predict the complex nonlinear temporal dynamics of various systems with memory [22,23]. Unlike typical recurrent neural networks (RNNs), for RC only the output weights are trained, making it significantly simpler and faster to train than most other RNNs and guaranteed to converge: RC does not suffer from the vanishing or exploding gradient problem [22,24]. Furthermore, RC does not require backpropagation, which is computationally intensive.

In this Article, we present results on spatially-distributed bulk optical RC using laser speckle in a multimode optical waveguide as the "reservoir", which allows for parallelized input and could potentially be scaled into a photonic integrated circuit (PIC) to form a multi-GHz-bandwidth, analog photonic processor. We show that the system can map the dynamics of a near-chaotic Mackey-Glass (MG) waveform. We demonstrate the optical system performed, with good accuracy, a multi-variate, multi-category classification task with preprocessed audio data.

The existing implementation of photonic RC can be broadly split into two categories: delay based [14-16,18-21] and optical node array based [7,8,9]. Although impressive, high-speed performance has been achieved with delay-based RCs, they are intrinsically limited by the total delay time, which is often on the order of a hundred nanoseconds [19, 21], which sets the limitation for real-time processing speeds and for chip-scale integration.

Similarly, the previous work by Dong et al. used free-space propagation and speckle generated from scattering from a rough surface [7,8]. This generates a larger number of neurons, but is also not compatible with high-speed, chip-scale integration. Here, we build up on the approach by Scofield et al. [9], which uses speckles generated by propagation through optical waveguides and a 1D array of light modulators. With this approach, we demonstrate that only 200 neurons are required to perform multivariate classification tasks. The essential components of our RC, an array of high-speed modulators and detectors, can be readily fabricated using commercial foundries with bandwidth exceeding 45 GHz [25]. In addition, the calculations presented here demonstrate that a 100-micron wide, 10-cm long planar waveguide provides sufficient speckle mixing for building a 100-neuron reservoir computer. Such planar waveguides can be fabricated in tight spirals using a silicon-on-insulator (SOI) PIC with a total footprint less than 1 cm$^2$ [25, 26]. This provides an immediate path for developing a scalable chip-scale RC system. The processing speed of such a device would only be limited by the optical delay of the chip, the modulator bandwidth, and the photodiode response time. This opens the possibility of performing real-time signal processing of radio-frequency (RF) and other high frequency signals, or rapid serial processing of very large datasets.

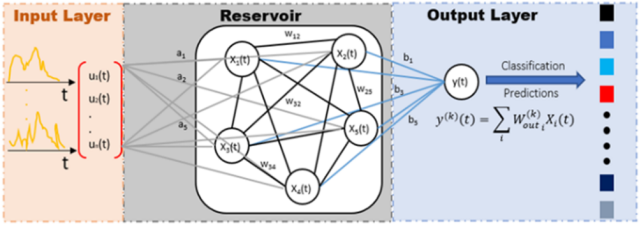

Figure 1: Topology of the reservoir computer consisting of a single input and a single output layer connected via a complex reservoir of neurons with random but fixed connections.

## Architecture and Experimental Demonstration of RC

A reservoir computer consists of three layers (Fig. 1): an input layer, which may accommodate multiple simultaneous inputs, a large reservoir of neurons with random but fixed connections, and an output layer, which may yield multiple outputs (e.g. prediction, classification). The reservoir effectively expands the dimensionality of the problem by distributing the input into multiple neurons. The recurrent nature of the reservoir computer makes it suitable for performing time-series prediction and classification.

The discrete time evolution of the neurons of a reservoir computer can be expressed as a recursive non-linear mapping of the form [22,24]:

$$X(t+1) = \alpha F_{NL}[W.X(t) + V.u(t)] + (1-\alpha)X(t) \quad (1)$$

where the N-dimensional column vector, $X(t)$ corresponds to the neuron values at time $t$, $\alpha \in (0, 1)$ is the leaking rate, $F_{NL}$ is a sigmoidal function (e.g. the logistic function, a hyperbolic tangent), $W$ is a matrix representing the connections between the neurons in the reservoir, $V$ is a matrix of the random weights between the inputs and the neurons, and $u(t)$ is the multi-dimensional input data used to train and test the reservoir computer.

In our physical implementation of a reservoir computer, the input data and feedback neuron values, weighted by leaking rate $\alpha$, are injected into the multimode fiber by intensity modulating the spatial profile of the light using a 1D array spatial light modulate (SLM). A multimode fiber transforms the input light and approximates the random weighted matrix operators $W$ and $V$. Non-linear activation is a critical step for machine-learning algorithms that maps between the input and the output response of the neurons. In our experiments this is achieved by using the non-linear saturation effect of the camera pixels $F_{NLcam}$, which performs an operation $F_{NL}[g(t)] = F_{NLcam}(|g(t)|^2)$ on the raw neuron values.

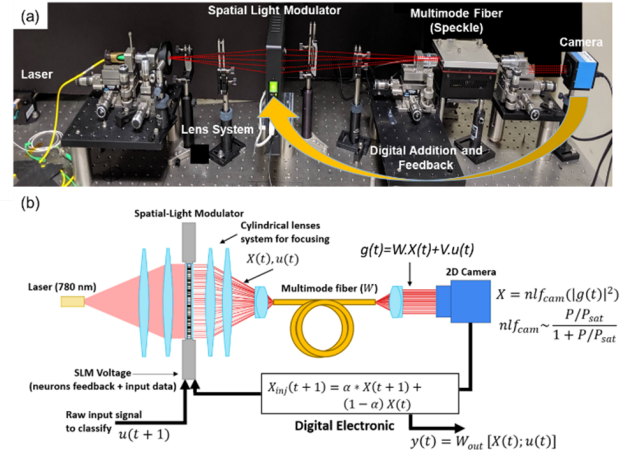

Figure 2: (a) Laboratory realization of speckle based optical reservoir hardware. The output of a 780 nm continuous-wave laser is modulated using a 1D SLM array and is imaged into a multi-mode fiber. The speckle pattern generated at the output of the multimode fiber is imaged using a 2D camera array, stored, added with the previous iteration of the experiment, and fed back into the SLM. (b) Schematic of the optical RC showing the optical and electronic processing part. The non-linear activation is performed through the saturation effect of the camera.

The laboratory implementation of the RC along with the schematic of the experiment is shown in Fig. 2 (a) and (b). A continuous-wave (CW) frequency stabilized single mode laser operating at 780 nm illuminates a 1D 12-bit SLM with 640 pixels using a pair of cylindrical lenses. The voltages of the SLM pixels are calibrated to optimize the modulation depth for the given laser wavelength. 200 pixels of the SLM are imaged onto the input of a 2.9 m long, 300-μm diameter, 0.39-NA multimode fiber such that each pixel excites different values of the modal expansion coefficients in the fiber. The propagation of several thousand optical modes through the fiber performs the equivalent of the random weight operations (multiplication by $W$ and $V$) needed for RC. The output of the fiber is imaged onto a 2D CCD camera, and the intensities of 200 pixels, corresponding to the neuron outputs of the reservoir computer, are recorded on a computer.

The neuron values times $\alpha$ are summed with the previous neuron values multiplied by $(1-\alpha)$. The neuron values are scaled to drive the SLM and address the SLM for the next timestep. The waveforms to be processed are also fed simultaneously to the SLM pixels. Multi-dimensional data can be fed into the waveguide by simultaneously modulating multiple additional SLM pixels. The output of the reservoir computer at time t is given by $y(t) = W_{out} X(t)$, where $W_{out}$ is a matrix of trained output weights. The state $X(t)$ of the reservoir neurons at each timestep is measured by running the system with the training time-series dataset input and collected into one state matrix $X = (X^T(t_1), X^T(t_2), ..., X^T(t_M))$, where $X \in \mathbb{R}^{N \times M}$ and $M$ is the total time steps, and then the single layer of output weights, $W_{out}$, for each classification task is calculated using regularized linear regression:

$$W_{out} = (XX^T + \lambda I_N)^{-1} XY, \quad (2)$$

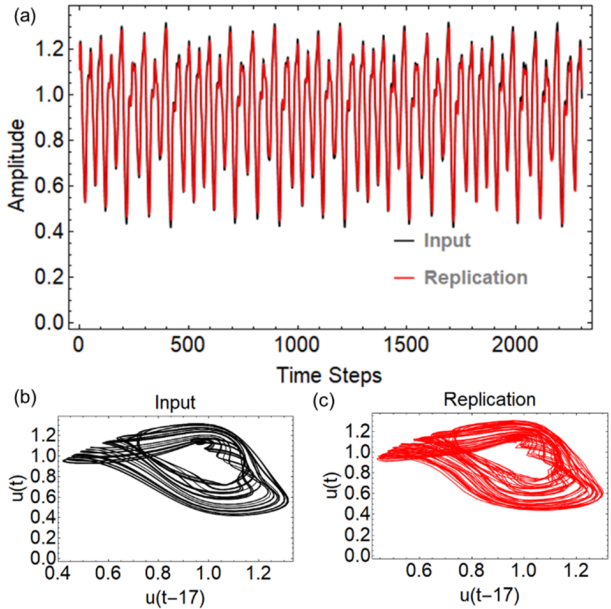

Figure 3: (a) Amplitude of the Mackey-glass function with τ =17as a function of time input to the optical RC (blue) and recovered using the trained output weights (red). Chaotic attractor for the (b) input and (c) recovered generated from the time delayed MG input and output from the RC. The plots show that the system operates in a high-dimensional regime and the RC is fully able to map the near-chaotic dynamics.

where $Y =(y(t_1), y(t_2), y(t_3),...)$ is the target output matrix, $I_N$ is the identity matrix, and $\lambda$ is the regularization parameter. In contrast to a tradition RNN, where all weights can be adjusted to train the network, training in RC is done for the output layer, which is typically fast, and is always convergent [20,24].

## Experimental Results

### Time Series Recovery

To demonstrate that the implemented RC hardware is capable of mapping complex near-chaotic waveforms into the reservoir and can replicate such a signal, we performed a time-series recovery experiment using the numerical solution to the MG equation with time delay (τ) of 17 [27]. The MG equation is a nonlinear time-delay differential equation that can generate complex, chaotic dynamics that are often found in naturally occurring systems with some temporal memory [27]. A 2300 timestep long MG solution time series is fed into the RC hardware and the output neurons are recorded. The first half of the neuron data (X) is used to train the output weights ($W_{out}$) such that the input waveform is recovered, and this calculated $W_{out}$ is used with the second half of the data to predict the input data. The input and recovered waveforms are plotted in Fig. 3(a). The chaotic attractors for the input and recovered MG timeseries are plotted in Fig. 3(b) and (c) respectively. The plots show that the system operates in a high- dimensional regime and the reservoir computer is fully able to map the near-chaotic dynamics with minimal noise. Since $W_{out}$ is calculated from the training dataset and then used to recover the test input data, the optical setup needs to be highly stable for the entire time it takes to run the training and test data in order to replicate such near-chaotic dynamics.

### Audio Classification

To demonstrate the ability of the experimental implementation of the optical reservoir computer to perform classification tasks [12,13,16, 20], we used the Japanese vowels public data set [28, 29]. This is a standard benchmark data set that consists of multivariate time-series data for a multiple-category classification task. The data consists of the first twelve Mel-frequency cepstrum coefficients (MFCCs) of utterances of the Japanese diphthong 'ae' by nine different speakers, with the task being to classify each utterance by speaker. 200 utterances, each padded with zeros to be 29 time-steps long, were used for both training and testing, with the testing utterances being different than those used for training.

The twelve MFCCs were fed in as inputs to the optical RC using twelve pixels of the SLM. Figure 4(a) shows a schematic of the experiment. As the input is fed in, the states of the individual neurons in the reservoir change (Fig. 4(b)). The state of the reservoir is collected at each time step, and the training portion is used to calculate $W_{out}$ using Eq. 2. This matrix is then used to predict the output and classify each utterance during the testing portion. The predicted output closely matches the input values (Fig. 4(c)), showing that the reservoir can reconstruct the input data accurately. This $W_{out}$ was also used to predict the classifications of the testing inputs. The predicted classification of each utterance was defined as the mode of the classification of each timestep within an utterance.

Figure 5(a) shows the predicted and correct classifications of the speakers for the training and testing utterances. The accuracies for the training and testing data sets are 88.5% and 81.5% respectively. The classification accuracy is expected to be higher for the training data set as $W_{out}$ was calculated from the neuron values corresponding to this data set. The system was still able to achieve very high classification accuracy for the testing data set, which consisted of 200 utterances the system had never seen before. The classification accuracy (81.5%) is much higher than that of random classification (11.1%). Using simulated RC with comparable number of neurons, we were able to obtain up to 98% classification accuracy. The reduced accuracy in the experiment is attributed to small thermal and mechanical instabilities of the multimode fiber. We expect these instabilities to be eliminated in a temperature controlled, packaged chip-scale device.

Speaker classification is not a trivial task as different utterances by the same speaker are not necessarily highly correlated and may in fact have large variations. Figure 5(b) shows the calculated correlation between each of the utterances in the full training data set (30 utterances per speaker). For some speakers (e.g. Speakers 5 and 6), different utterances by the same speaker are highly correlated. However, for other speakers (e.g. Speaker 3), different utterances by the same speaker show very little correlation. If one were to attempt classification based on correlation alone, it would be very difficult to distinguish highly correlated but distinct speakers, such as Speakers 8 and 9.

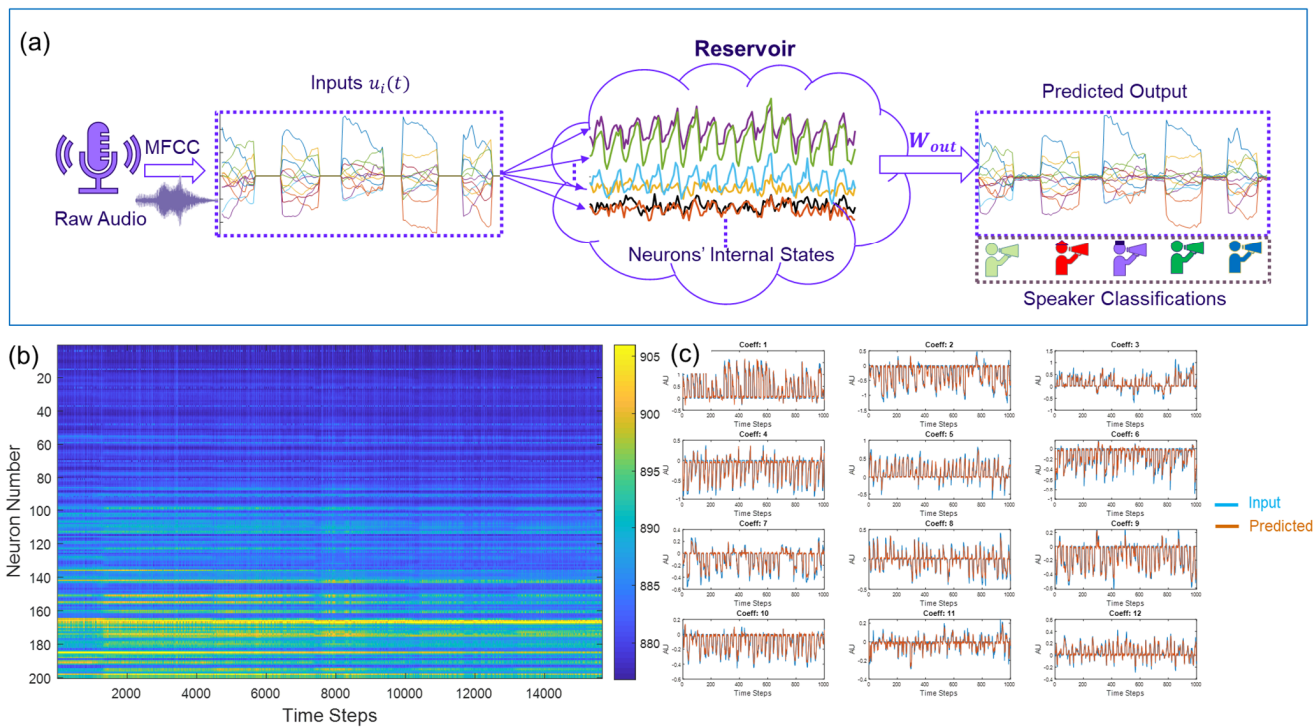

Figure 4: (a) A high-level schematic of the experiment with the measured data. The raw audio signals are digitally converted into their MFCCs. The first 12 MFCCS are used as the input *u(t)* into the reservoir, changing the states of the individual neurons in the reservoir. At each time step, the matrix of the neurons states is multiplied by $W_{out}$ in order to get the output and the predicted speaker classifications. (b) is experimentally recorded value of each neuron as a function of time. (c) shows the predicted value (orange) of each of the 12 MFCCs calculated from the output of the reservoir, compared to the input value (blue) of the MFCCs.

Figures 5(c) and 5(d) show confusion plots for the same training and testing data sets as in Figure 5 (a). The data is normalized such that each column, corresponding to each correct speaker, adds up to 100%. The diagonal elements show the percentage of correctly classified utterances per speaker, and the off-diagonal elements show the percentage of incorrectly classified utterances. These plots show that the utterances by certain speakers (e.g. Speaker 8) are more likely to be misclassified, whereas utterances by certain speakers (e.g. Speaker 6) were correctly classified with very high accuracy. These confusion plots also show which classification errors are more likely (e.g. Speaker 1 is likely to be misidentified as Speaker 8, but never Speaker 6).

The results reported here are for a complicated multivariate data set; for simple header classification problems, we achieve 100% classification accuracy.

## Simulations

### *Reservoir Size, Leaking Parameter, and Noise*

A simulated reservoir computer was also trained and tested with the Japanese vowels data set for comparison to the performance of the experimental implementation. For the simulation, $V$ was a randomly generated matrix with values uniformly distributed between -0.5 to 0.5, and $W$ was a randomly generated matrix with values from -0.5 to 0.5 and then normalized such that the spectral radius was 1.25. Hyperbolic tangent was chosen as the non-linear function $F_{NL}$. $W_{out}$ was then calculated by training the simulated reservoir computer with the training portion of the Japanese vowels data set. This $W_{out}$ was used to predict the classifications of the testing portion of the Japanese vowels data set. Each utterance is zero-padded as necessary to be 29 time steps long, and the predicted classification for each utterance is determined by a "winner-take-all" algorithm such that the label predicted for 15 or more time steps is taken to the label for the utterance. The accuracy was then determined by comparing the predicted classifications to the correct classifications.

Figure 6(a) shows the accuracy of the test data set classification as a function of leaking rate and reservoir size. Even with a small reservoir size (25 neurons), we were able to achieve >90% accuracy. The accuracy was not very sensitive to the leaking rate for leaking rates >0.2. For a detailed analysis on the performance comparison between RC and traditional RNN, see Ref. [24].

To test the reservoir computer's robustness to noise, we added Gaussian noise to the test data. Figure 6(b) shows the classification accuracy as a function of the amplitude of the Gaussian noise for two cases. The classification accuracy drops as the amplitude of the noise increases. Furthermore, increasing the number of neurons does not improve the robustness to noise. The achieved classification accuracy with our experimental hardware is therefore not limited by the number of neurons, but by noise and other non-idealities in the system.

This simulation study demonstrates that a few hundred neurons are sufficient to perform complex classification tasks with very high accuracy. Such numbers of neurons can be implemented by fabricating an array of high-speed modulators and detectors on a chip-scale device, using readily available standard foundry processes [25].

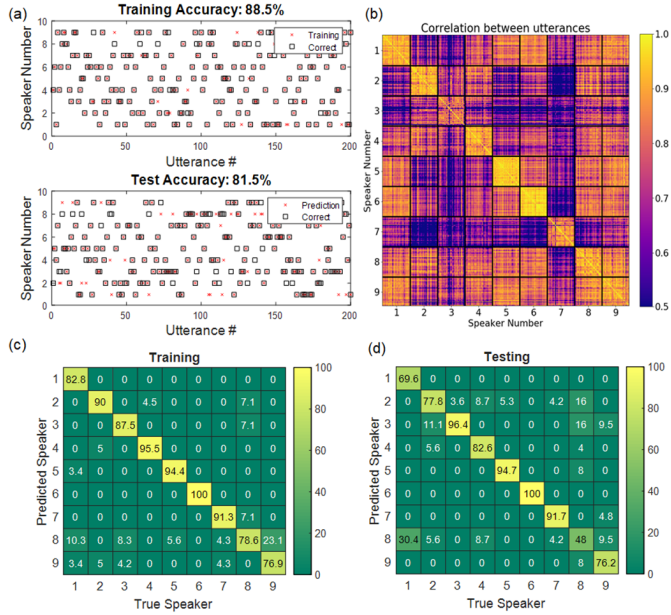

Fig. 5. (a) Predicted (red crosses) vs correct (black boxes) classification of the speakers for the training (top) and test (bottom) dataset. Overlap between a cross and box indicates that the utterance was correctly classified. (b) calculated cross-correlation between each of the training utterances. The lighter yellow signifies highly correlated sounds and purple less correlation. (c) and (d) show confusion plots for the same training and testing data sets as in Figure (a).

### Simulation of Planar Waveguides for PIC Realization of RC

Our experiments and those of Dong et al. [8,9] show empirically that speckle in multimode fiber and speckle from rough surface scattering provide excellent random projections for RC. For chip-scale integration of the spatially-distributed RC architecture, it is important to demonstrate that the random matrix generation and multiplication can be performed in a planar multimode waveguide. Here we perform detailed full wave calculations that demonstrate speckle in a multimode planar waveguide also works extremely well for RC and gives performance equivalent to that obtained by using a fully numerical simulation based on projections with a pseudo-random matrix with entries uniformly distributed between -1 and +1.

We start with the well-known solutions for the TM modes $\Psi_i(x, z)$ in planar dielectric waveguides [30] of the form

$$\Psi_i(x,z) = [A_i \cos(h_i x) + B_i \sin(h_i x)] \exp(i \beta_i z) \qquad (3)$$

where $x$ is the transverse dimension and $z$ is the direction of propagation, $A_i$ and $B_i$ are constants determined by the boundary conditions and input field, $h_i$ is the transverse mode wavenumber, and $\beta_i$ is the propagation constant. We ignore the field in the cladding as this is unimportant for the speckle pattern used in RC and use solutions for the TM mode. The parameters $h_i$ and $\beta_i$ are obtained by numerical solution of the standard transcendental equation for waveguide modes [30,31]. For a 100-micron wide silicon on $SiO_2$ waveguide as shown in Fig. 7(e), solution of this transcendental equation yields about 500 roots each with its own spatial pattern. These modes are excited at 100 1-micron-wide input ports, they mix as they propagate down the 10-cm long waveguide, and finally they form the speckle pattern to be measured at 100 output ports, also 1 micron wide. Since the propagation constants $\beta_i$ are different for each mode, the interference pattern of all the modes varies rapidly with propagation down the guide. If one input port is excited, less than 5 cm of propagation is needed for this 1-micron input field to fill the guide.

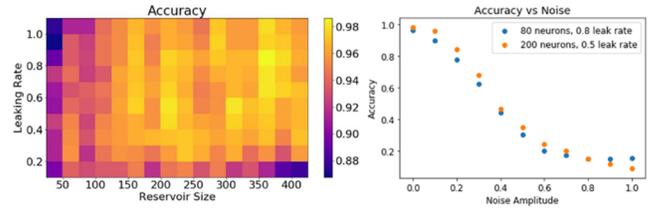

Figure 6: Reservoir computer simulated on a computer using random matrices. (a) Accuracy of the test data set classification as a function of leaking rate and reservoir size. (b) Classification accuracy as a function of amplitude of Gaussian noise added to the test data for (blue) 80 neurons and 0.8 leak rate, and (orange) 200 neurons and 0.5 leak rate.

The first step in the calculations is to find the expansion coefficients for each of the 100 input fields. These expansion coefficients are multiplied by the values of the appropriate feedback neuron [elements of the vector $X(t)$ in Section 2] or the input signal $u(t)$. Second, the electric field at the output of the guide is calculated by finding the superposition of all these modes. Next, we take the absolute value squared of the field at the output plane and integrate it over the individual output pixels. Then a nonlinearity is applied to these values and they are added to a fraction (1-α) of the neurons at the previous time step. We tried simulations using only the square law non-linearity but these perform poorly. In all our calculations we need an additional sigmoidal nonlinearity. Experimentally, we find that the square-law and saturation response of the camera pixels is sufficient.

To summarize the full wave calculations, the matrix products $W.X$ and $V.u$ in Eq. (1) are replaced by modulation of the modal expansion coefficients at the input to the guide, propagation down the guide and calculation of the intensity of the optical speckle field at the output of the guide. Fig. 7 (a-d) compare the results for the MG waveforms using speckle mixing as described above to a conventional calculation based on Eq. (1) with a random matrix $W$ that is calculated from random numbers uniformly distributed between -1 and +1. The top row gives the test MG waveform superposed on the recovered MG waveform and the difference between the recovered and test waveforms for the conventional calculation while the bottom row shows the same task using the full wave speckle mixing calculation. The parameters of the calculations are 99 neurons, leaking rate $\alpha$ = 0.5 and nonlinearity $F_{NL}$ = 1/[1+exp(-x)]. The training sequence consisted of 8 Mackey-Glass waveforms with time delays from 6 to 20 in steps of 2. The test sequence consisted of the same 8 MG waveforms but in random order. The obvious conclusion is that for MG waveforms the planar-waveguide speckle-based reservoir computer performs nearly identically to a reservoir computer in which a random matrix is used. We have performed similar calculations for test and training waveforms composed of multiple sinusoids—again with no significant difference.

We note that both the speckle calculations take a lot longer than the random matrix calculations, but of course, there is no intent to make a numerical reservoir computer based on speckle calculations. The intent is to do the speckle "calculations" passively using the optical RC hardware in the time it takes light to propagate through the waveguide.

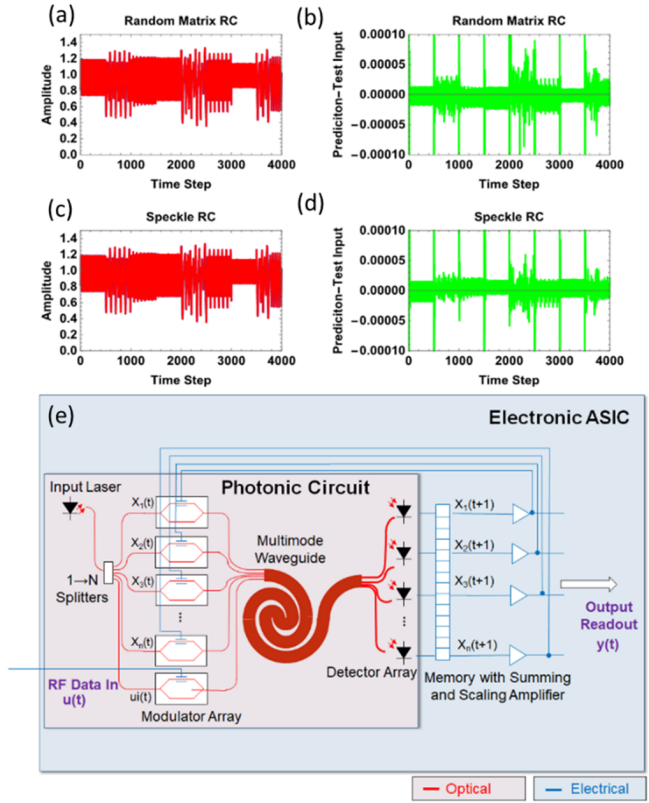

Figure 7: (a) Mackey-Glass amplitude recovery and test. Amplitude of recovered signal (red) and input signal (blue) as a function of time step for a random matrix reservoir computer. (b) Difference between recovered and input signals for random matrix. (c) Amplitudes for speckle-based RC. (d) Difference for speckle-based reservoir computers. (e) Concept schematic for a chip-scale reservoir computer consisting of a single laser, an array of modulators, a multimode planar waveguide, and an electronic adder and amplifier to drive the modulator with the analog signal. The shaded red region consists of optical components and the blue region consists of electrical components.

In addition to the time-series reproduction, we perform a classification task using the full wave simulator. The performance of the speckle-based RC was nearly identical to that of numerical RC. These results demonstrate the viability of a speckled-based RC architecture implemented in a chip-scale platform using a planar multimode waveguide.

Simulations also show that multiple small reservoir computers (less than ~100 neurons) can be used simultaneously with the same input waveform to increase the effective number of neurons, and this implies that large reservoir computers (~1000s of neurons) can be constructed from multiple copies of the same small PIC. Such multiple near-identical copies of the PIC can be readily fabricated through a commercial foundry.

An example schematic of the integrated circuit implementation of the reservoir computer is given in Fig. 7(e). The 1D SLM can be replaced with an array of on-chip modulators, the optical fiber can be implemented as a planar multimode waveguide, and the camera can be replaced with an array of on-chip integrated detectors [25]. The laser can also be fabricated on-chip with indium phosphide technology or surface mounted on the $SiO_2$ platform. It is important to note that the intention is not to implement all-optical RC, as a hybrid device can take advantage of the strengths of both optical and electrical systems to achieve a multi-GHz, low size, weight, and power hardware accelerator. For example, using the detectors, the non-linearity operation can be performed with no added processing complexity. The summing and scaling of the signal to drive back the modulators can be performed electrically at multi-GHz clock rate with a simple application-specific integrated circuit (ASIC) that can be seamlessly integrated with the photonic chip.

## Conclusions

In this Article, we demonstrated an optical reservoir computer using speckles generated from a multimode waveguide. We demonstrated that the hardware can perform a classification task of multivariate time series audio data. We performed full wave optical calculations of the speckle-based optical reservoir computer implemented in a chip-scale platform with a 100 μm wide $SiO_2$ multimode planar-waveguide and demonstrated that the system performed as well as a fully numerical implementation for time-series reconstruction and classification tasks. The RC architecture demonstrated here through bulk optics can be readily built on an integrated platform combining a photonic chip fabricated at a commercial foundry with electrical ASIC components. The demonstrated architecture and results provide a pathway for building a hybrid hardware accelerator capable of performing on the order of GHz bandwidth real-time signal processing.

**Funding.** The work is funded by an Internal Research and Development (IR&D) grant from The Aerospace Corporation.

**Disclosures.** The authors declare no conflicts of interest